\documentclass[pdflatex,sn-mathphys-num]{sn-jnl}

\usepackage{graphicx}%
\usepackage{caption}
\usepackage{subcaption}
\usepackage{subcaption, calc}
\usepackage{adjustbox}
\usepackage{graphicx}
\usepackage{multirow}%
\usepackage{ragged2e}%
\usepackage{amsmath,amssymb,amsfonts}%
\usepackage{amsthm}%
\usepackage{mathrsfs}%
\usepackage[title]{appendix}%
\usepackage{xcolor}%
\usepackage{textcomp}%
\usepackage{manyfoot}%
\usepackage{booktabs}%
\usepackage{algorithm}%
\usepackage{algorithmicx}%
\usepackage{algpseudocode}%
\usepackage{listings}%
\usepackage{physics}%
\usepackage{multirow}
\usepackage[per-mode=symbol]{siunitx}  
\usepackage{chemformula}  
\usepackage{orcidlink}
\usepackage[capitalise]{cleveref}  

\expandafter\def\expandafter\normalsize\expandafter{%
    \normalsize%
    \setlength\abovedisplayskip{12pt}%
    \setlength\belowdisplayskip{12pt}%
    \setlength\abovedisplayshortskip{12pt}%
    \setlength\belowdisplayshortskip{12pt}%
}

\newcommand{\mumaxp}{mumax$^+$} 
\newcommand{\mumaxthree}{mumax$^3$}

\newcommand{\GitHub}{GitHub}

\newcommand{\tensor}[1]{\hat{\vb{#1}}}  

\begin{document}

\title[ \mumaxp{}: extensible GPU-accelerated micromagnetics and beyond]{ \mumaxp{}: extensible GPU-accelerated micromagnetics and beyond}


\author[1]{\fnm{Lars} \sur{Moreels}\orcidlink{0000-0001-7362-2466}}
\author[1]{\fnm{Ian} \sur{Lateur}\orcidlink{0009-0003-8283-4083}}
\author[1]{\fnm{Diego} \spfx{De} \sur{Gusem}\orcidlink{0009-0008-3341-0555}}
\author[1]{\fnm{Jeroen} \sur{Mulkers}\orcidlink{0000-0002-4000-9275}}
\author[1]{\fnm{Jonathan} \sur{Maes}\orcidlink{0000-0003-4973-8383}}
\author[2]{\fnm{Milorad V.} \sur{Milo\v{s}evi\'c\orcidlink{0000-0002-5431-377X}}}

\author*[1]{\fnm{Jonathan} \sur{Leliaert}\orcidlink{0000-0001-8778-3092}}
\author[1]{\fnm{Bartel} \spfx{Van} \sur{Waeyenberge}\orcidlink{0000-0001-7523-1661}}

\affil[1]{DyNaMat, \orgdiv{Department of Solid State Sciences}, \orgname{Ghent University}, \postcode{9000} \city{Ghent}, \country{Belgium}}
\affil[2]{NANOlab Center of Excellence \& Department of Physics, \orgname{University of Antwerp}, \postcode{2020} \city{Antwerp}, \country{Belgium}}

\abstract{
%
%

We present \mumaxp{}, an extensible GPU-accelerated micromagnetic simulator with a Python user interface, to address the challenges posed by current magnetism research into systems with complex magnetic ordering and interfaces. It is a general solver for the space- and time-dependent evolution of the magnetization and related vector quantities, using finite difference discretization.
Here, we present its application and design and discuss features not available in \mumaxthree{}, such as the modeling of antiferromagnets with magnetoelastic coupling. As an illustration of its capabilities, we use \mumaxp{} to simulate state of the art magnetic systems. Specifically, we demonstrate the current induced domain wall motion in a polycrystalline antiferromagnet, we simulate the working principle of a strain-driven antiferromagnetic racetrack memory and we reproduce experimentally observed domain structures in a non-collinear antiferromagnet.
}




\maketitle
{\small
\noindent\textbf{*Corresponding author:}\\ 
\href{mailto:jonathan.leliaert@ugent.be}{jonathan.leliaert@ugent.be} 
}

\section{Introduction}

Micromagnetic calculations have been an indispensable tool for the development of the field of modern magnetism. Especially the availability of open-source~\cite{oommf,mumax1,mumax2,magpar,mumax3,Nmag, magnumfe} and GPU-accelerated software packages~\cite{Fastmag,Kakay2010,mumax3, Boris, micromagnum, magnumfd} boosted research by enabling efficient numerical studies of nano- and micrometer-sized magnetic systems, complementary to experimental investigations.
More recently, research efforts have shifted to more complex magnetic systems such as antiferromagnets~\cite{RevModPhys.90.015005} and 2D magnetic heterostructures (Van der Waals systems)~\cite{geim_van_2013}. Different ways of manipulating the magnetization are being explored as well, e.g. by applying strain~\cite{bandyopadhyay_perspective_2024} or electric fields~\cite{bandyopadhyay_magnetic_2021}.

In order to model these modern materials, software tools must go beyond the standard micromagnetic theory; preferably, they should be easily extensible to include other physical phenomena of interest. With this in mind, we designed \mumaxp{}, a versatile GPU-accelerated micromagnetic simulation package. It calculates the space- and time-evolution of nano- to microscale magnets using a finite difference scheme. The program has been designed independently of \mumaxthree{}~\cite{mumax3} and features a Python~\cite{python} based user interface. This allows users to directly use common packages such as NumPy~\cite{numpy}, SciPy~\cite{scipy} and Matplotlib~\cite{matplotlib} to process and view their data directly. The Python module wraps the main part of the code, which is written in C++~\cite{C++} and CUDA~\cite{nvidia}. The latter allows calculations to be performed in parallel on a GPU, reducing simulation time compared to CPU-based codes.

The extensible framework of \mumaxp{} is in stark contrast to the design of \mumaxthree{}. This extensibility provides a plethora of new possibilities, many of which are brought together in the implementation of (non-collinear) antiferromagnets with full support for magnetoelastic interactions. Therefore, in this paper we will demonstrate the application of these two features. It will place emphasis on the practical and contemporary importance in the research field of modern magnetism.

We examine the domain wall motion in a polycrystalline antiferromagnet under the influence of an applied current. Furthermore, we show the working principle of a strain-driven magnetic racetrack in \ch{NiO} and, lastly, reproduce nitrogen-vacancy magnetometry images of the non-collinear antiferromagnet \ch{Mn3Sn}.
In the final section, we provide a brief description of the design of the software package along with an overview of the implemented physics.\\


The development of \mumaxp{} is motivated by the ``plus'' mindset, i.e. versatility and extensibility, combined with user-friendliness, in favor of performance and speed. Our intention is not to replace \mumaxthree{}, but rather to offer a platform where additional functionalities can be easily developed and used.

All capabilities of the code have been thoroughly tested against the micromagnetic standard problems~\cite{mumag}, \mumaxthree{}~\cite{mumax3} and analytical results. These tests will not be repeated here, but can be found freely in the \GitHub{} repository (\url{https://github.com/mumax/plus}), along with the complete open-source code under the GPLv3 license. Details about dependencies, installation instructions and the API can also be found in the repository.

\section{Demonstrations}
This section is devoted to the showcasing of the implementation of antiferromagnets and elastodynamics, which is not present in \mumaxthree{} \cite{mumax3, tutorial3}. More details about the used methods and implementations, as well as some key differences with \mumaxthree{}, are explained in section \ref{sec:methods}.

\subsection{Current-driven domain wall mobility in a polycrystalline antiferromagnetic strip}
\label{ssec:afm_mobility}

\begin{figure}[h]
    \centering
    \begin{subfigure}[t]{0.49\textwidth}
        \makebox[0pt][l]{\textbf{(a)}\hspace{1em}}%
        \hspace{2em}%
        \adjustbox{valign=t}{\includegraphics[width=\linewidth-2em]{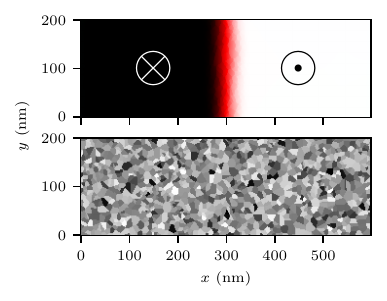}}
        \phantomsubcaption
        \label{fig:polycryst_afm}
    \end{subfigure}
    \hfill
    \begin{subfigure}[t]{0.49\textwidth}
        \makebox[0pt][l]{\textbf{(b)}\hspace{1em}}%
        \hspace{2em}%
        \adjustbox{valign=t}{\includegraphics[width=\linewidth-2em]{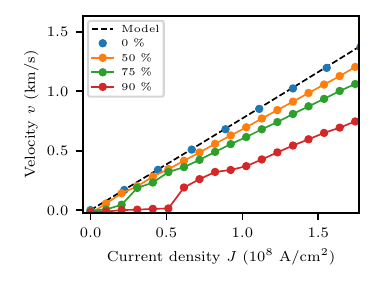}}
        \phantomsubcaption
        \label{fig:mobility_data}
    \end{subfigure}
    \caption{Simulation setup \textbf{(a)} and mobility curves \textbf{(b)} of a polycrystalline antiferromagnet. Each sublattice has a uniaxial out-of-plane anisotropy with a strength of \SI{64}{\kilo\joule\per\meter\cubed}, a saturation magnetization of \SI{400}{\kilo\ampere\per\metre}, interfacial DMI strength of \SI{0.11}{\milli\joule\per\meter\squared} and a damping constant of 0.1 \cite{Giovanni}. The homogeneous antiferromagnetic exchange constant is set at \SI{-50}{\pico\joule\per\meter}.\\
    The top panel in \textbf{(a)} shows the relaxed N\'eel vector of a polycrystalline antiferromagnetic strip. The domain wall, in which the N\'eel vector lies along the positive $x$ direction, separates two out-of-plane domains. The ferromagnetic and inhomogeneous antiferromagnetic exchange constants, \SI{10}{\pico\joule\per\meter} and \SI{-10}{\pico\joule\per\meter} respectively, are reduced by $75\,\%$ at the grain boundaries.
    The bottom panel in \textbf{(a)} shows the different grains inside the material, generated using a Voronoi tessellation \cite{JL_mobility} with an average grain diameter of \SI{10}{\nano\meter}.\\
    Panel \textbf{(b)} shows the simulated domain wall velocity in an antiferromagnetic strip under the influence of an applied current density, with each curve corresponding to a different percentage reduction in exchange constants at the grain boundaries. The dashed curve shows the analytical model derived in Ref. \cite{Giovanni} for the monocrystalline case.}
    \label{fig:afm_mob}
\end{figure}

Understanding the mobility properties of domain walls in various materials has been an important field of study for the development of future spintronic devices \cite{DW_logic, AFM_spintronics, ncafm_cluster_2020, JL_DW_mob}. We can use \mumaxp{} to investigate such dynamics in antiferromagnets. Here, we study the current-induced motion of a domain wall in an antiferromagnetic strip with polycrystalline structure. The domain wall is stabilized by interfacial DMI and can be moved by applying a spin-polarized current in the direction of the strip, effectively causing a spin transfer torque as formulated by Slonczewski \cite{Slonczewski}. Following \mumaxthree{}, we are able to define multi-grain geometries inside a material by generating a Voronoi tessellation. The ferromagnetic and antiferromagnetic exchange constants are then reduced at the different grain boundaries \cite{JL_mobility}.\\
We simulate a strip with a width of \SI{200}{\nano\metre} and a thickness of \SI{10}{\nano\metre}. The strip is discretized into cells of size 2$\cross$2$\cross$\SI{10}{\nano\metre\cubed} and we define grains with an average diameter of \SI{10}{\nano\metre}. This simulation setup is illustrated in \cref{fig:polycryst_afm} and described in more detail in the caption of \cref{fig:afm_mob}.

We monitor the velocity of the domain wall in a time frame of \SI{0.5}{\nano\second} for different values of the current density and for different reductions of the exchange constants at the various grain boundaries. Results can be found in \cref{fig:mobility_data}. We see a distinct decrease in the slope of the velocity curves for decreasing values of the intergrain exchange along with a stronger domain wall pinning at lower current densities. These results are very similar to those reported in Refs. \cite{JL_mobility, JL_pinning}, where similar simulations based on permalloy have been performed. The dashed curve in \cref{fig:mobility_data} represents the model derived in Ref. \cite{Giovanni} for the ideal case without grains. Our simulation data for a $0\,\%$ reduction of the exchange constants, effectively disregarding any grain boundaries, agree very well with this model.

\subsection{Strain-driven antiferromagnetic racetrack memory}\label{sec:racetrack}
Previous studies have demonstrated the possibility to move domain walls in both ferromagnets and antiferromagnets through the application of surface acoustic waves or strain gradients \cite{AFM-strain-DW-motion, multiferroic-strain-DW-motion, strain-DW-motion-gradient}.
Building on this concept, we use \mumaxp{} to investigate the controlled movement of domain walls in antiferromagnetic \ch{NiO} under the influence of strain caused by the application of traction.
The simulation consists of 512$\cross$128$\cross$1 cubic cells with a side length of \SI{3}{\nano\metre}. The magnet has an easy anisotropy axis along the length axis of the magnet. This stabilizes two out-of-plane domain walls, separating domains in which the N\'eel vector lies along the easy axis. Four small notches are introduced
from \SI{300}{\nano\metre} onward with \SI{150}{\nano\metre} spacing, which serve as pinning sites for the domain walls. The material parameters used in the simulation can be found in \cref{tab:racetrack_parameters}.

\begin{figure}[h]
    \centering
    \includegraphics[width=\textwidth]{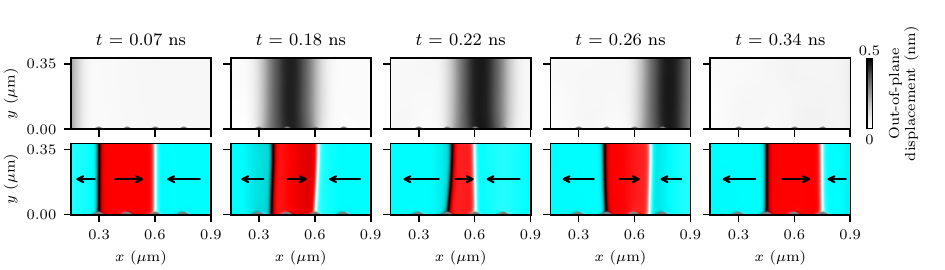}
    \caption{
    An elastic wave is generated through the application of a traction pulse at the left edge at \SI{0}{\micro\metre}, and subsequently
    propagates through the antiferromagnetic \ch{NiO} film. This wave mobilizes the domain walls, which are initially pinned at the notches at \SI{300}{\nano\metre} and \SI{600}{\nano\metre}, as discussed in \cref{sec:racetrack}.
    The top row shows the out-of-plane elastic displacement, while the bottom row shows the N\'eel vector of the antiferromagnet.
    The in-plane domains, oriented in the directions indicated by the arrows, are separated by out-of-plane domain walls, shown in black or white, whose N\'eel vectors lie into, and out of the plane of the film, respectively.
    Different snapshots after the introduction of the pulse at $t=\SI{0}{\nano\second}$ are shown. The domain walls have moved to the second and fourth notches at \SI{450}{\nano\metre} and \SI{750}{\nano\metre} respectively, after passage of the pulse.}
    
    
    \label{fig:elastic_racetrack}
\end{figure}

Initially two domain walls are placed on the first and third notch. In order to move both domain walls one notch along the strip, we 
generate an elastic out-of-plane shear pulse by applying one period of a sinusoidal out-of-plane shear traction wave with a frequency of \SI{8}{\giga\hertz} and an amplitude of \SI{0.3}{\giga\pascal} at a boundary. The resulting dynamics are shown in \cref{fig:elastic_racetrack}. The induced strain, approximately \SI{0.14}{\percent} in amplitude, detaches the domain wall from the first notch via the Villari effect~\cite{ferromagnet-physics} and transports it along the length axis of the magnet. The elastic pulse, traveling with a velocity of about \SI{4}{\kilo\metre\per\second}, overtakes the domain wall which gets pinned at the second notch. The pulse then detaches the second domain wall which is transported to the final notch in the same manner.
This effectively simulates the underlying working principle of a traction-mediated magnetic racetrack \cite{racetrack}. This concept can be extended to various domains with varying sizes in a single antiferromagnetic strip where each domain corresponds to a series of 0 or 1 bits. The domain walls can be shifted one notch at a time by the controlled application of traction pulses.

\begin{table}[h]
    \centering
    \caption{The \ch{NiO} simulation parameters used for the strain-driven domain wall motion discussed in Section \ref{sec:racetrack}, which apply to both sublattices. The magnetic parameters are based on Refs. \cite{NiO-exchange,NiO-params,NiO-damping}, with the exception of the interfacial DMI, which is chosen to stabilize the domain wall. The (magneto)elastic parameters are taken from Ref. \cite{NiO-elastic}, aside from the stiffness coefficient of the Rayleigh damping. The latter is chosen to yield a damping ratio of about \SI{5}{\percent} for shear waves with a wavelength of \SI{24}{\nano\metre}.}
    \label{tab:racetrack_parameters}
    \begin{tabular}{|p{3.1cm}|c||p{3.1cm}|c|}
        \hline
        \multicolumn{2}{|c||}{\textbf{Magnetic parameters}} & \multicolumn{2}{c|}{\textbf{(Magneto)elastic parameters}} \\
        \hline
        Saturation \newline magnetization $M_{\rm S}$ & \SI{425}{\kilo\ampere\per\metre} & First magnetoelatic \newline coupling constant $B_1$ & \SI{-15}{\mega\joule\per\metre\cubed}  \\
        Ferromagnetic exchange constant $A_{11}$ & $\sim\,$\SI{4.4}{\pico\joule\per\metre} & Second magnetoelastic \newline coupling constant $B_2$ & \SI{-8.5}{\mega\joule\per\metre\cubed} \\
        Homogeneous antiferromagnetic exchange $A_0$ & $\sim\,$\SI{-46}{\pico\joule\per\metre} & Mass density $\rho$ & \SI{6853}{\kilogram\per\metre\cubed} \\
        Lattice constant $a$ & \SI{0.42}{\nano\metre} & Stiffness tensor \newline component $C_{11}$ & \SI{330}{\giga\pascal} \\
        Uniaxial anisotropy & \SI{85.7}{\kilo\joule\per\metre\cubed} & Stiffness tensor \newline component $C_{12}$ & \SI{60}{\giga\pascal} \\
        Gilbert damping $\alpha$ & \num{2.1d-4} & Stiffness tensor \newline component $C_{44}$ & \SI{110}{\giga\pascal} \\
        Interfacial DMI & \SI{0.7}{\milli\joule\per\metre\squared} & Stiffness coefficient of \newline the Rayleigh damping& \SI{0.1}{\pico\second} \\
        \hline
    \end{tabular}
\end{table}

\subsection{Stray field imaging of the non-collinear antiferromagnet \ch{Mn3Sn}}
\label{ssec:ncafm}

The non-collinear antiferromagnet \ch{Mn3Sn} is a prototypical material in the research field of antiferromagnetic spintronics \cite{Mn3Sn_spintronics, NCAFM_STT, Mn3Sn_exc, Mn3Sn_anisu}. Owing to the interplay between exchange, DMI and magnetocrystalline anisotropy, it exhibits a small net magnetic moment \cite{Mn3Sn_anisu, NCAFM_STT}, which makes direct imaging of magnetic structures more evident than in collinear antiferromagnets \cite{NCAFM_imaging, NCAFM_NV, Mn3Sn_MOKE}.

\begin{figure}[h]
     \centering
    \begin{subfigure}[t]{0.47\textwidth}
        \makebox[0pt][l]{\textbf{(a)}\hspace{1em}}%
        \hspace{2em}%
        \adjustbox{valign=t}{\includegraphics[width=\linewidth-2em]{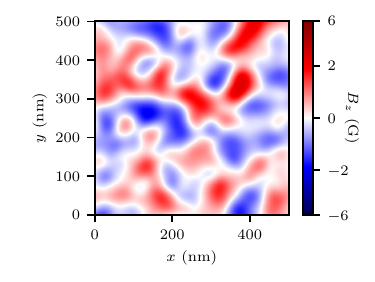}}
        \phantomsubcaption
        \label{fig:ncafm_sim}
    \end{subfigure}
     \hfill
    \begin{subfigure}[t]{0.51\textwidth}
        \makebox[0pt][l]{\textbf{(b)}\hspace{1em}}%
        \hspace{2em}%
        \adjustbox{valign=t}{\includegraphics[width=\linewidth-2em]{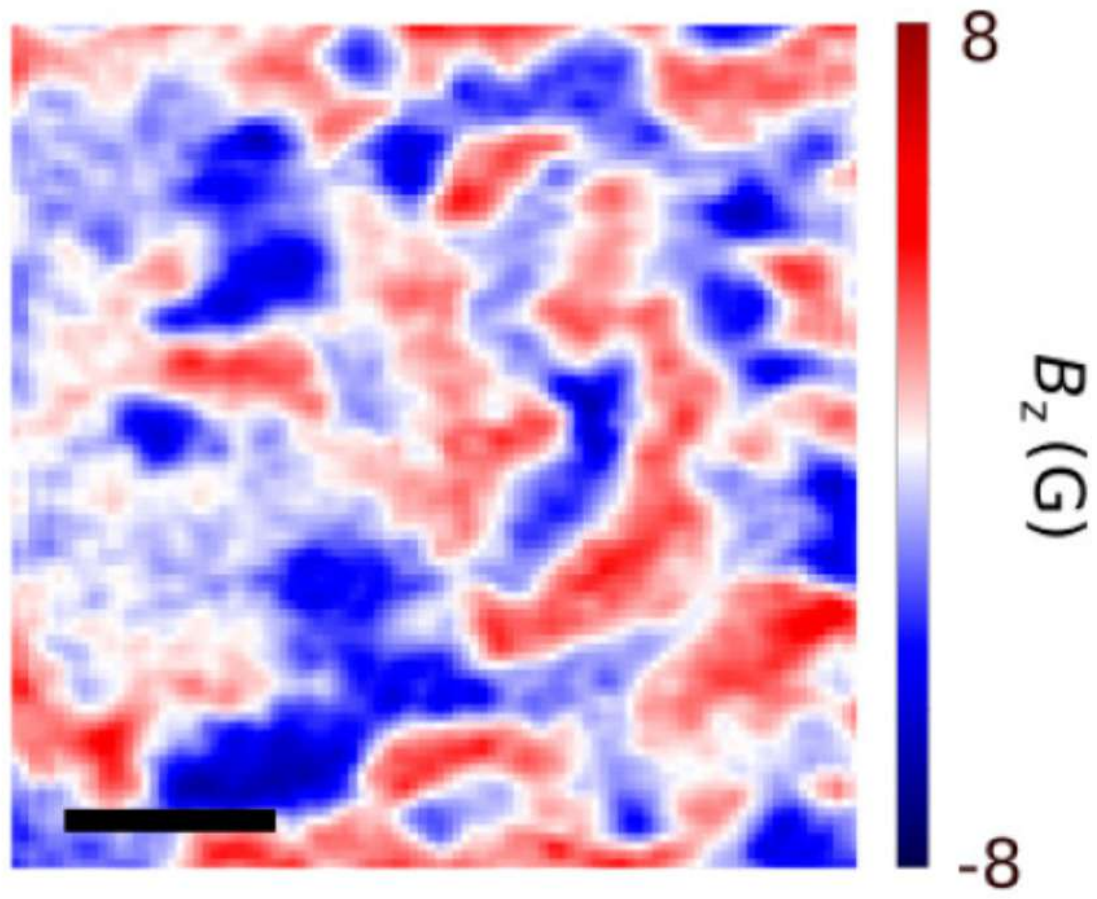}}
        \phantomsubcaption
        \label{fig:ncafm_exp}
    \end{subfigure}
     \caption{The out-of-plane component of the stray field of a \ch{Mn3Sn} thin film, as discussed in section \ref{ssec:ncafm}.\\
     Panel \textbf{(a)} shows the calculated field using \mumaxp{} after letting a non-collinear antiferromagnet relax from a random state. Each sublattice has a saturation magnetization of \SI{1}{\mega\ampere\per\meter}, uniaxial anisotropy constant of \SI{5}{M\joule\per\meter\squared} (based on Ref. \cite{NCAFM_STT}), a Gilbert damping constant of 0.01 and a typical lattice constant of \SI{3.5}{\angstrom}. The ferromagnetic, inhomogeneous and homogeneous antiferromagnetic exchange constants are \SI{10}{\pico\joule\per\meter}, \SI{-15}{\pico\joule\per\meter} and \SI{-25}{\pico\joule\per\meter} respectively. The homogeneous DMI vector is set at $\SI{10}{\mega\joule\per\meter\cubed}$ in the direction perpendicular to the film, while the interfacial DMI has a value of $\SI{3.5}{\milli\joule\per\meter\squared}$ \cite{NCAFM_NV}.\\
     Panel \textbf{(b)} shows an NV measurement image of a \ch{Mn3Sn} thin film with a thickness of \SI{30}{\nano\metre}, measured $\sim\SI{60}{\nano\metre}$ above the film's surface. The scale bar denotes \SI{0.5}{\micro\metre}. Reprinted (adapted) with permission from Li, S. et al. (2023). Nanoscale Magnetic Domains in Polycrystalline Mn3Sn Films Imaged by a Scanning Single-Spin Magnetometer. Nano Letters, 23(11), 5326–5333. \url{https://doi.org/10.1021/acs.nanolett.3c01523}. Copyright 2023 American Chemical Society.}
     \label{fig:ncafm}
\end{figure}

Li et al. \cite{NCAFM_NV} imaged magnetic domain structures using nitrogen-vacancy (NV) magnetometry measurements of \ch{Mn3Sn} films. Here, we demonstrate the simulation of the measured magnetostatic field using \mumaxp{} by considering an antiferromagnet with three distinct exchange-coupled sublattices. We define a square 500$\cross$500$\cross$\SI{30}{\nano\metre\cubed} non-collinear antiferromagnetic film discretized into cells of size 1$\cross$1$\cross$\SI{30}{\nano\metre\cubed}. Grains with an average diameter of \SI{40}{\nano\metre} are generated using a Voronoi tessellation \cite{JL_mobility}. Each grain has an easy axis which lies either out-of-plane, perpendicular to the film's surface, or in a random in-plane direction \cite{NCAFM_NV}. Furthermore, the exchange coupling between neighboring grains is reduced to $10\%$ of the intragrain value. To reproduce the NV scans shown in Ref. \cite{NCAFM_NV}, we calculate the out-of-plane component of the magnet's stray field \SI{60}{\nano\metre} above the film's surface. The results, along with the simulation parameters, can be found in \cref{fig:ncafm}. The agreement between Figures \ref{fig:ncafm_sim} and \ref{fig:ncafm_exp}, along with the order of magnitude of the stray field showcases the ability of \mumaxp{} to simulate non-collinear antiferromagnetic systems.

\section{Methods}
\label{sec:methods}

The \mumaxp{} module is written in C++ and CUDA for GPU support and can be compiled using either single or double precision. The module is wrapped in Python and can be imported in an input script like any other package. The object-oriented character of Python and C++ allows for a very flexible and transparent use of classes and accompanying instances. For example, each magnetic instance has their own set of material parameters and multiple magnets from different magnet classes can simultaneously coexist. This contrasts with \mumaxthree{}, where the user has to resort to defining a limited number of \emph{regions} in order to create inhomogeneities in material parameters \cite{mumax3}. Other advantages of the use of a Python-based interface are prominent, e.g. the use of NumPy's \cite{numpy} intuitive array indexing logic to set material parameters in different subsets of cells. These benefits are illustrated in the various examples in the \href{https://github.com/mumax/plus}{\GitHub{}} repository.

In the remainder of this section we outline the physics included in \mumaxp{} with an emphasis on the differences compared to \mumaxthree{}. A more detailed description of the different functionalities and features can be found in the API.

\subsection{Micromagnetism}

For each magnet defined in the simulation world, \mumaxp{} solves the dynamic equation
\begin{equation}
    \pdv{\vb{m}}{t} = \boldsymbol{\tau}\,,
    \label{eq:magnetic-dynamic-equation}
\end{equation}
where $\vb{m}$ is the magnetization normalized with respect to the saturation magnetization $M_\text{S}$ and $\boldsymbol{\tau}$ is the sum of the Landau-Lifshitz-Gilbert (LLG) torque \cite{LLG} and any additional spin-transfer torques as formulated by Zhang and Li~\cite{Zhang-Li} or Slonczewski~\cite{Slonczewski}, which are implemented in the same way as in \mumaxthree{}~\cite{mumax3}. Several embedded Runge-Kutta methods with the option to use adaptive time stepping \cite{Heun, Bogacki, Cash, Fehlberg, DORMAND} are available for time integration. A list can be found in \cref{tab:solvers}.\\
The LLG torque is given by
\begin{equation}
    \boldsymbol{\tau}_{\text{LLG}} = -\frac{\gamma}{1+\alpha^2}\left[\, \vb{m}\cross\vb{B}_\text{eff} + \alpha\left(\vb{m}\cross\left(\vb{m}\cross\vb{B}_\text{eff}\right)\right)\, \right]\,,
    \label{eq:LLG_torque}
\end{equation}
with $\gamma$ the gyromagnetic ratio and $\alpha$ the dimensionless Gilbert damping constant~\cite{LLG,Gilbert}.
The different field terms which play a role and add up to the effective field $\vb{B}_\text{eff}$ are calculated based on the current magnetization state $\vb{m} \equiv \vb{m}(\vb{r},t)$. These fields include an external field, the uniaxial or cubic anisotropy field, the exchange field, a demagnetizing field, stray fields from other magnets, the field induced by the Dzyaloshinskii-Moriya interaction, a thermal field and the magnetoelastic field. Of these, the external field, anisotropy fields and thermal field \cite{stochasticLLG} are implemented in the same way as in \mumaxthree{}. Their details will not be repeated here,
and we refer the interested reader to their thorough description in Ref.~\cite{mumax3}.

The remainder of this section highlights new features which are not present in \mumaxthree{}. While our focus is on ferromagnets and collinear antiferromagnets, the underlying formalism naturally generalizes to ferrimagnets and non-collinear antiferromagnets. The former is achieved by considering two different sublattices in an antiferromagnet, while the latter can be done by adding a third sublattice with the same interaction terms as the first two.

\begin{table}[ht]
\centering
\caption{Runge-Kutta solvers in \mumaxp{} together with their convergence order and error estimate order which is used for adaptive time stepping.}
\begin{tabular}{|c|c|c|}
 \hline
 \multirow{ 2}{*}{Method} & Convergence & Error\\
 & order & estimate order\\
 \hline
 Heun (RK12) & 2 & 1 \\ 
 Bogacki-Shampine (RK23) & 3 & 2 \\
 Cash-Karp & 5 & 4\\
 Fehlberg (RKF45) & 5 & 4\\
 Dormand-Prince (RK45) & 5 & 4\\
 \hline
\end{tabular}\label{tab:solvers}
\end{table}

\subsubsection{Exchange field}
The exchange field present in an antiferromagnet acting on sublattice $s$ under the influence of the other sublattice $s'$ is described in Ref.~\cite{Giovanni} and given by
\begin{equation}
    \vb{B}_{\text{exch}}^{(s)} = \frac{2A_{11}^{(s)}}{M_\text{S}^{(s)}} \laplacian \vb{m}^{(s)} +
                              \frac{A_{12}}{M_\text{S}^{(s)}} \laplacian \vb{m}^{(s')} +
                              \frac{4A_0}{M_\text{S}^{(s)} a^2}\vb{m}^{(s')}\,.
    \label{eq:AFM_exchange_field}
\end{equation}
The first term in \cref{eq:AFM_exchange_field} accounts for the standard ferromagnetic exchange interaction. This is the only term present when simulating ferromagnetic systems and is implemented as in \mumaxthree{} \cite{mumax3}. The second term contains the inhomogeneous exchange stiffness constant $A_{12} <$ \SI{0}{\joule\per\metre} and describes a similar antiferromagnetic exchange interaction between sublattices. The last term accounts for the homogeneous exchange interaction between antiferromagnetically coupled spins within the same simulation cell. This interaction is described by a homogeneous exchange constant $A_0 <$ \SI{0}{\joule\per\metre} and a lattice parameter $a$.

\subsubsection{Magnetostatic field}
The calculation of the magnetostatic field is similar to the method used in OOMMF \cite{oommf}. It is done by using the analytical result \cite{Newell-demag} below a certain cutoff radius and the asymptotic result \cite{asymptotic-demag} above that radius. In the case of antiferromagnets, the vector sum of the spins within a simulation cell is computed prior to evaluating the resulting field.

\subsubsection{Dzyaloshinskii–Moriya interaction and boundary conditions}
In \mumaxp{}, the user can set the 9 independent components of the antisymmetric rank-$3$ tensor $\tensor{D}$ ($D_{ijk}= -D_{ikj}$) of the Dzyaloshinskii–Moriya interaction (DMI), which encompasses the DMI strengths and chiral properties of the considered lattice. 
Since the DMI tensor is determined by the crystallographic and magnetic symmetries of the material, each sublattice in an antiferromagnet can have its own crystallographic environment and thus its own DMI tensor. Furthermore, a third DMI tensor is used to describe the intersublattice interaction. Generally, the energy density due to DMI in an antiferromagnet can be written as~\cite{AFM_DMI_BC,General_AFM_DMI}
\begin{equation}
    \mathcal{E} = \sum_{s,s'}\sum_{i,j,k} D_{ijk}^{(s, s')} m_j^{(s)} \partial_im_k^{(s')}\,,
    \label{eq:dmi}
\end{equation}
where the indices $(i,j,k)$ are spatial, $(s, s')$ denote the different sublattices and it is assumed that $D_{ijk}^{(s, s')} = D_{ijk}^{(s', s)}$. The ferromagnetic expression is obtained by considering $s=s'$.

The tensorial character of the DMI also comes into play when calculating the Neumann boundary conditions. Given the structure of the exchange field in Eq.~\eqref{eq:AFM_exchange_field}, the boundary conditions for a sublattice $s$ in an antiferromagnet are given by~\cite{AFM_DMI_BC,Giovanni}
\begin{equation}
    2 A_{11}^{(s)} \partial_{\vb{n}}\vb{m}^{(s)} = A_{12} \left(\partial_{\vb{n}} \vb{m}^{(s')} \cross\vb{m}^{(s)} \right)\cross\vb{m}^{(s)} + \tensor{D}^{(s)}\cdot\left(\vb{n}\otimes\vb{m}^{(s)}\right)\,,
    \label{eq:AFM_Neumann_BC}
\end{equation}
where $\vb{n}$ is the surface normal and a term proportional to the inhomogeneous exchange constant $A_{12}$ takes the interaction with the other sublattice into account. This term is not present when considering ferromagnets.\\

The Neumann boundary conditions as given in Eq.~\eqref{eq:AFM_Neumann_BC} are the default in \mumaxp{}. Alternatively, open or periodic boundary conditions can be used. The chosen boundary conditions will not only affect the DMI field calculations, but also the ferromagnetic and antiferromagnetic inhomogeneous exchange calculations. Periodic boundary conditions also affect the calculation of the magnetostatic fields.\\

Apart from the inhomogeneous interaction described by \cref{eq:dmi}, \mumaxp{} also considers a homogeneous DMI between spins within the same simulation cell, given by the energy density \cite{DMI_homo_skyr, DMI_homo_fast}
\begin{equation}
    \mathcal{E} = \vb{d}\cdot\left(\vb{m}^{(s)}\cross\vb{m}^{(s')}\right)\,,
\end{equation}
where $s$ and $s'$ denote two distinct sublattices and $\vb{d}$ is the DMI vector. The latter lies parallel to the principal axis of the material \cite{DMI_homo_skyr}.

\subsection{Elasticity}
Apart from the magnetodynamic equation (\ref{eq:magnetic-dynamic-equation}), \mumaxp{} can also solve the second order elastodynamic equation for cubic crystal symmetries~\cite{wave-motion-solids, wave-propagation-solids}
\begin{equation}
    \rho \frac{\partial^2 \vb{u}}{\partial t^2} = \vb{f}_{\text{tot}}\,,
    \label{eq:elasticity}
\end{equation}
with $\vb{u}$ the elastic displacement, $\rho$ the mass density and $\vb{f}_\text{tot}$ the total body force density acting on the magnet.
%
The implementation is mostly based on a previous extension of \mumaxthree{}~\cite{finite-diff-magnetoelastic-sim}, except that the elastic force $\vb{f}_{\text{el}}$ does not use second order differences of the elastic displacement.
Rather, $\vb{f}_{\text{el}}$ is computed as the numeric divergence of the stress $\tensor{\sigma}$, while the strain is calculated with the numeric gradient of the displacement.
In this way, traction and traction-free boundary conditions~\cite{barber-elasticity} were easily implemented. These impose $\tensor{\sigma} \cdot \vb{n} = \vb{t}$ at the boundary, where $\vb{n}$ is the normal to the boundary and $\vb{t}$ is the applied traction.
Both spatial derivative calculations utilize a fourth order accurate central difference scheme~\cite{numerical-analysis} in the bulk material. Lower order schemes are used close to and at the boundaries.

A second difference between the \mumaxthree{} extension described in Ref. \cite{finite-diff-magnetoelastic-sim} and our implementation, is the addition of viscous damping, where an extra term, proportional to the strain rate, is added to the total stress tensor \cite{2000-mechanical-response-polymers, note-generalization-Kelvin-Voigt-model, mechanical-behavior-materials}. This tends to dampen high frequency oscillations \cite{rheological-interpretation-rayleigh-damping}.

\subsection{Magnetoelasticity}\label{sec:magnetoelasticity}

Magnetoelasticity for ferromagnets with cubic crystal symmetry is implemented in the same way as described in Ref.~\cite{finite-diff-magnetoelastic-sim}. It is the combination of magnetostriction~\cite{ferromagnet-physics} and the Villari effect or inverse magnetostriction~\cite{ferromagnet-physics}. In the former the magnetization affects the elastic displacement by applying a magnetoelastic force given by
\begin{equation}
    \vb{f}_{\text{mel}} = 2 B_1 \mqty[
        m_x \partial_x m_x \\ m_y  \partial_y m_y \\ m_z \partial_z m_z
    ]
    + B_2 \mqty[
        m_x \left( \partial_y m_y + \partial_z m_z \right) + m_y \partial_y m_x + m_z \partial_z m_x \\
        m_y \left( \partial_x m_x +  \partial_z m_z \right) + m_x \partial_x m_y + m_z \partial_z m_y \\
        m_z \left( \partial_x m_x + \partial_y m_y \right) + m_x \partial_x m_z + m_y \partial_y m_z \\
    ]\,
    ,
    \label{eq:magnetoelastic-force}
\end{equation}
while in the Villari effect the elastic strain affects the magnetization by generating an effective field given by
\begin{equation}
    \vb{B}_{\text{mel}} = - \frac{2}{M_\text{S}} \left( B_1 \mqty[
        \varepsilon_{xx} m_x \\ \varepsilon_{yy} m_y \\ \varepsilon_{zz} m _z
    ]
    + B_2 \mqty[
        \varepsilon_{xy} m_y + \varepsilon_{xz} m_z \\
        \varepsilon_{xy} m_x + \varepsilon_{yz} m_z \\
        \varepsilon_{xz} m_x + \varepsilon_{yz} m_y \\
    ]
    \right)\,.
    \label{eq:magnetoelastic-field}
\end{equation}
In both \cref{eq:magnetoelastic-force} and \cref{eq:magnetoelastic-field} $B_1$ and $B_2$ are the magnetoelastic coupling constants and $\tensor{\varepsilon}$ is the strain tensor.

For antiferromagnets, magnetoelasticity is implemented similarly to how it is reported in Ref.~\cite{AFM-near-THz}. Both sublattices have a shared elastic displacement, velocity, stiffness, strain and body force, while each ferromagnetic sublattice $s$ has its own magnetoelastic coupling constants $B_1^{(s)}$ and $B_2^{(s)}$, magnetization $\vb{m}^{(s)}$ and saturation magnetization $M_\text{S}^{(s)}$. Each sublattice then creates a magnetoelastic body force $\vb{f}_{\text{mel}}^{(s)}$, which is added to the total body force. Conversely, each sublattice magnetization is affected by a unique magnetoelastic field $\vb{B}_{\text{mel}}^{(s)}$ created by its own magnetization $\vb{m}^{(s)}$ and the shared strain.




\section{Conclusions and outlook}
As a demonstration of the capabilities of \mumaxp{} that are not present in \mumaxthree{}, we first presented the current-driven domain wall mobility in a polycrystalline antiferromagnet. Furthermore, the addition of magnetoelastics has provided a way to simulate the working principle of a racetrack memory in \ch{NiO} driven by strain through the application of traction. Finally, we were able to reproduce nitrogen-vacancy magnetometry images in the non-collinear antiferromagnet \ch{Mn3Sn}.

We presented the micromagnetic model implemented in \mumaxp{} and highlighted the key differences from its predecessor \mumaxthree{}. These advancements include a Python-based interface, support for calculations involving (non-collinear) antiferromagnets and ferrimagnets, and the integration of magnetoelasticity across all magnet types in the package.


The development of \mumaxp{} is ongoing and more functionalities are expected to be added in future releases. Some possibilities include adaptive time stepping appropriate for magnetoelastic simulations, a more general stiffness tensor and piezoelectric coupling.
Since \mumaxp{} is open-source and built with an object-oriented design for easy extensibility, anyone can fork the repository (\url{https://github.com/mumax/plus}) and enhance the module with additional features.

\section*{Acknowledgments}
We acknowledge financial support from the SHAPEme project (EOS ID 400077525) from the FWO and F.R.S.-FNRS under the Excellence of Science (EOS) program.

J.L. was supported by the Fonds Wetenschappelijk Onderzoek (FWO-Vlaanderen) with postdoctoral fellowship No. 12W7622N.

We would like to thank Oleh Kozynets for his software engineering additions to the code.


\bibliography{sn-bibliography}

\end{document}